\documentclass[11pt]{article}
\usepackage{times}
\usepackage{graphicx}
\usepackage{epsfig}
\usepackage{amsmath}

\advance \topmargin by -\headheight
\advance \topmargin by -\headsep

\evensidemargin \oddsidemargin
\newcommand{\qed}{\hfill \rule{2mm}{3mm}}
\newenvironment{proof}{\par \noindent{\bf Proof:}}{\(\qed\) \par}
\newcommand{\eat}[1]{}

\newcommand{\hide}[1]{}
\newcommand{\junk}[1]{}

\newtheorem{theorem}{Theorem}
\newtheorem{lemma}{Lemma}[theorem]

\newtheorem{defn}{Definition}[theorem]

\newtheorem{observation}[theorem]{Observation}

\newcounter{ccc}
\newcommand{\bcc}{\setcounter{ccc}{1}\theccc.}
\newcommand{\icc}{\addtocounter{ccc}{1}\theccc.}
\newcommand{\wa}{{\cal W}}
\newcommand{\wai}{{\cal W}^{-1}}
\newcommand{\zr}{{\cal Z}_R}
\newcommand{\real}{{\cal R}}

\title{How far will you walk to find your shortcut: 
Space Efficient Synopsis Construction Algorithms}
\author{Sudipto Guha \thanks{Department of Computer Science, University of Pennsylvania, 3330 Walnut St, Philadelphia, PA 19104. Email: {\tt sudipto@cis.upenn.edu}}}

\begin{document}
\date{}
\maketitle
\begin{abstract}
  In this paper we consider the wavelet synopsis construction problem
  without the restriction that we only choose a subset of coefficients
  of the original data. We provide the first near optimal algorithm.
  
  We arrive at the above algorithm by considering space efficient
  algorithms for the restricted version of the problem. In this context 
  we improve previous algorithms by almost a linear factor and reduce the 
  required space to almost linear. Our techniques also extend to histogram
  construction, and improve the space-running time tradeoffs for V-Opt
  and range query histograms. We believe the idea applies to a broad
  range of dynamic programs and demonstrate it by showing improvements
  in a knapsack-like setting seen in construction of Extended Wavelets.
\end{abstract}
\section{Introduction}
Wavelet synopsis techniques have become extremely popular in query
optimization, approximate query answering and a large number decision
support systems. Wavelets, specially Haar wavelets, are one-one mappings and 
admit a natural 
multi-resolution interpretation, as well as 
fast algorithms for the forward and inverse transforms.
 
Given a set of $n$ numbers $X=x_1,\ldots,x_n$ the wavelet synopsis
construction problem seeks to choose a synopsis vector $Z$ with at
most $B$ non-zero entries, such that the inverse wavelet transform of
$Z$ (denoted by $\wai(Z)$) gives a good estimate of the data. The
typical objective measures are (suitably weighted\footnote{ The
  weighted $\ell_k$ with weights $\pi_i>0$ and wlog $\sum_i \pi_i=n$
  minimizes $\left( \sum_i (\pi_i (x_i -
    \wai(Z)_i))^k\right)^{\frac1k}$.}) $\ell_k$ norm of $X - \wai(Z)$.
In an early paper \cite{MVW98}, demonstrated a number of different
applications for wavelet synopsis and proposed greedy algorithms.
However for objective measures other than the $\ell_2$ measure,
the greedy algorithm does not necessarily provide the optimum solution.  The
problem is quite non-trivial, primarily due the fact that the Wavelet
basis vectors overlap and cancellations (subtractions) occur. This
means that we can have two coefficients that cancel out each other
leaving a significantly (exponentially) smaller contribution, which
needs to be accounted for.  The precision of the coefficients in the
optimum solution can be much larger than the precision of the data. In
fact there are no known bounds or promising techniques for quantifying
the precision - this is the biggest stumbling block in the synopsis
construction.

Most of the literature focuses on the {\bf Restricted case} where the non-zero
entries of $Z$ are equal to the corresponding entries in the
transform of the original data, $\wa(X)$. A natural question remains: {\em why
  should we be optimizing under the restriction of retaining the
  coefficients of the data -- with no guarantees that such a
  restriction does not compromise the quality of the final synopsis?}
This is clearly suboptimal -- a comparable example would be to
optimize the synopsis for point queries, and use it for range queries.

A simple example renders the discussion concrete; $X=\{1,2,3,7\}$ and
$B=1$ illustrates that choosing any single coefficient of
$\wa(X)=\{3.25,-1.75,-0.5,-2\}$ (non-normalized) 
does not give the optimum answer for
$\ell_1$ or $\ell_\infty$ norm. 
Normalization does not help. The normalized transform is 
$\{4.55,-2.45,-0.5,-2\}$ -- but choosing the first coefficient as 
$4.55$ in the normalized setting implies assigning $4.55/\sqrt{2}=3.25$ everywhere. 
Thus dynamic program approaches that seek to see the effect of the coefficient on the data come to the same conclusion in both settings.
The optimum choices of $Z$ are
$\{z,0,0,0\}$ for any $2\leq z\leq 3$ and $\{4,0,0,0\}$ for $\ell_1$
and $\ell_\infty$ respectively. 
The same example applies to {\bf weighted} $\mathbf \ell_2$,
e.g., if $\pi=\{\frac12,\frac12,\frac32,\frac32\}$ then the best error achieved by retaining any single entry of $\wa(X)$ 
is $5.78$ whereas $Z=\{4.65,0,0,0\}$ gives an error of $4.87$.   
The example can be extended to any
$B$ (by repetition and scaling). The restriction of only retaining 
the coefficients of the data is significantly self
defeating.

However the restriction does ease the search for a solution, and as
this paper shows, is an important stepping stone towards the final
result. For the restricted case, \cite{GG02} gave a probabilistic
scheme (the space constraint is preserved in expectation only, along
with the error) and very recently \cite{GK04} gave an optimal solution.
This has been extended and improved in
\cite{muthu-wave}.  {\em However, the solution to the unrestricted
  case has remained elusive and we provide the first near optimal
  solutions.} In the process, we also improve upon previous algorithms
for the restricted case as well.  However our algorithm is best
explained by taking a different path, which brings us to the {\em major
theme} of the paper.

\paragraph{}
Synopsis construction is perhaps most relevant in context of massive
data sets. In some scenarios we can justify that the synopsis is
created using a ``scratch'' space larger than the synopsis and stored.  
However a {\em quadratic} or extremely superlinear space complexity is 
near infeasible for large $n$. 
The dependence on synopsis size $B$ is also important in this context --
the smaller the dependence is, the larger is the synopsis that can be 
computed in the environment of a particular system. Further, 
{\em space is typically a more inflexible resource}, 
and not just a matter of wait. However a natural conceptual 
question arises: {\em We are only given $n$ numbers, -- do we
  really need to save so much information to compute the optimum
  answer ?}

All previous algorithms (for the restricted case) are expensive in
space (see table below). This (super-linearity in $n,B$) is also seen
in context of histogram construction (we provide a detailed table in
Section~\ref{sec:hist}).  To avoid this expensive space complexity,
several researchers have introduced the notion of {\em working space},
which is the amount of space required to compute the error -- the rest
of the space is used to construct the answer (coefficients,
representatives, etc.).  In case of wavelets the working space used by
previous algorithms is $O(nB)$. In case of histograms, known
algorithms reconstruct the answer only using the $O(n)$ working space,
but {\em with a penalty of an extra factor of $B$ in the running
  time}.  In this paper, we reduce the space for wavelets and
eliminate the penalty for histograms, in fact our results show that
the {\em working space notion is not needed} for a wide range of
problems. To summarize {\bf Our contributions}:
\begin{itemize}\parskip=-0.05in
\item We provide the first near optimum algorithm for the wavelet
  synopsis construction problem. The algorithm naturally extends to
  multiple dimensions.
\item For the restricted case \cite{GG02} provided approximation algorithms, however the space constraints were obeyed in expectation. The results for (optimum) algorithms with strict space bounds are
\footnote{In \cite{muthu-wave} the space bounds are not explicitly provided, but
the total space appears to be $O(n^2B/\log B)$ as well. The authors of \cite{yossi1} consider the same problem for a
  non-Haar basis, and is excluded from the discussion here}:
\begin{center}
{\small
\begin{tabular}{|c|c|c|c|c|c|}
\hline
Paper &  Error & Time & Space & Working Space \\
\hline
\hline
\cite{GK04} & $\ell_\infty$ & $ O(n^2B \log B) $ & $O(n^2B)$ & $O(nB)$\\
         & $\ell_k$ & $ O(n^2B^2) $ & $O(n^2B)$ & $O(nB)$\\
\hline
\cite{muthu-wave} 

& (weighted) $\ell_k$ & $O(\frac{n^2B}{\log B}) $ & ? &  ? \\
\hline
\hline
{\bf This Paper} & (weighted) $\ell_\infty$& $O(n^2)$ & $O(n+B\log (n/B))=O(n)$ & $O(n+B\log (n/B))=O(n)$\\
& (weighted) $\ell_k$& $O(n^2 \log B)$ & $O(n)$ & $O(n)$\\
\hline
\end{tabular}
}
\end{center}
\cite{GK04} also provided approximation algorithms for multiple dimensions 
and our techniques extend to this context as well, and improves the
running time and space by almost a factor $B$. 

\item We improve several histogram construction algorithms, e.g.,
  V-Opt histograms, range query histograms, by simultaneously
  achieving the best known running time and space bounds. The results
  and {\em a table comparing the results} are presented in
  Section~\ref{sec:hist}. Due to lack of space omit the improvements
  for the range query histograms, which are similar.
\item We believe the space efficient paradigm is applicable to other
  dynamic programs as well, and we demonstrate the improvements in
  case of Extended Wavelets in Section~\ref{sec:ext}.
\end{itemize}

\section{The Restricted (Haar) Wavelet Synopsis construction Problem}
\label{sec:old}
\begin{figure}
\begin{center}
\begin{tabular}[t]{cc}
\begin{minipage}{2in}
\centerline{\psfig{figure=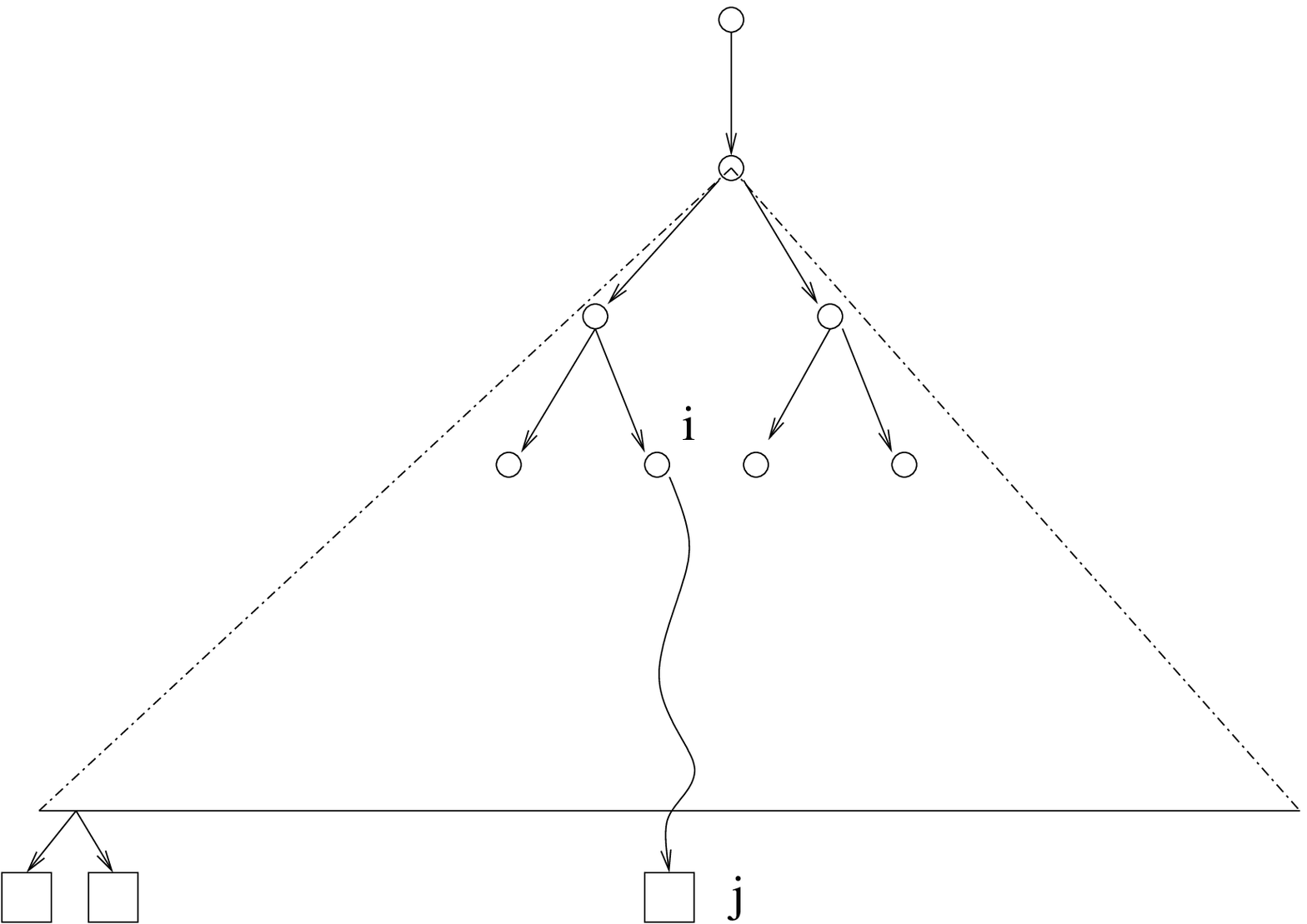,width=1.8in}}
\end{minipage}
& \framebox{
\small
\begin{minipage}{4in}
At each internal node $i$ to compute $E[i,b,S]$:
\begin{itemize}\parskip=-0.05in
\item We determine if we are choosing the coefficient $i$. 
\item Assuming we are, we decide how the remaining $b-1$ coefficients are to allocated between the two subtrees. If the children are $i_L$ and $i_R$,  we are interested in 
\vspace{-0.1in}
\[ \min_{b'} E[i_L,b',S\cup\{i\}]+ E[i_R,b-1-b',S\cup\{i\}] \]
\item Assuming that we do not choose $i$ we are interested in a similar expression giving the overall minimization to be 
\vspace{-0.1in}
\[
\min \left \{ \begin{array}{l} \min_{b'} E[i_L,b',S\cup\{i\}]+ E[i_R,b-1-b',S\cup\{i\}] \\ 
\min_{b'} E[i_L,b',S]+ E[i_R,b-b',S] \end{array} \right. 
\]
\end{itemize}
\end{minipage}
} \\
(a) & (b) 
\end{tabular}
\caption{The Error Tree and the previous algorithm. \label{fig:one} }
\end{center}
\end{figure}

We will work with {\bf non-normalized} wavelet transforms where the
inverse computation is simply adding the coefficients that affect a
coordinate\footnote{For normalized wavelets the normalization
  constant appears both in forward and inverse transform, all the
  results in the paper will carry over in that setting as well,with
  the introduction of the normalization constants at several places}.
The wavelet basis vectors are defined as (assume $n$ is a power of $2$):
\[ \begin{array}{rll}
  V_0(j) = & 1 & \mbox{~~for all $j$}\\
  V_{2^s+t}(j) = & \left\{ \begin{array}{l}
          1 \\
          -1 \end{array} \right. &
          \begin{array}{l}
          \mbox{for $(t-1)\frac{n}{2^s}+1 \leq j \leq \frac{tn}{2^s} - \frac{n}{2^{s+1}}$} \\
          \mbox{for $\frac{nt}{2^s}- \frac{n}{2^{s+1}}+1 \leq j \leq \frac{tn}{2^s} $} \\
\end{array} \hspace{0.5in} (1\leq t\leq \frac{n}{2^s}, 1\leq s \leq \log n)
\end{array}
\]
The above definitions ensure $\wai(Z) = \sum_i Z_i V_i$. To compute
$\wa(X)$, the algorithms computes the average
$\frac{x_{2i+1}+x_{2i+2}}{2}$ and the difference
$\frac{x_{2i+1}-x_{2i+2}}{2}$ for each pair of consecutive elements as
$i$ ranges over $0,2,4,6,\ldots$ The difference coefficients form the
last $n/2$ entries of $\wa(X)$. The process is repeated on the $n/2$
average coefficients - {\em their difference coefficients yield the
  $n/4+1,\ldots,n/2$'th coefficients of $\wa(Z)$}. The process stops
when we compute the overall average, which is the first element of
$\wa(Z)$. The wavelet basis functions naturally form a complete binary
tree since their support sets are nested and are of size powers of $2$
(with one additional node as a parent of the tree, see
Figure~\ref{fig:one}).  The $x_j$ correspond to the leaves, denoted by
boxes, and the coefficients correspond to the non-leaf nodes of the
tree.  This tree of coefficients is termed as the error tree
(following \cite{GK04}).  Likewise assigning a value $c_i$ to the
coefficient corresponds to assigning $+c_i$ to all leaves $j$ that are
{\bf left descendants} (descendants of the left child) and $-c_j$ to
all right descendants.  The leaves that are descendants of a
coefficient are termed as the {\bf support} of the coefficient.
Recall that the {\bf Restricted} (Haar) Wavelet construction problem
is that given a set of $n$ numbers $X=x_1,\ldots,x_n$ the problem
seeks to choose at most $B$ terms from the wavelet representation
$\wa(X)$ of $X$, say denoted by $\zr$, such that a (weighted) $\ell_k$
norm of $X - \wai(\zr)$ is minimized.

\subsection{Reviewing Previous Algorithm(s)}

It is immediate that the value of $\wai(\zr)_j$ is fixed by the choices of
all coefficients $i$ such that $j$ belongs to the support of $i$.
Suppose $S$ is a subset of the ancestors of a coefficient $i$.  Thus a
natural dynamic program emerges where we define $E[i,b,S]$ to be {\em
  the minimum contribution to the error from all $j$ in the support of $i$, such
  that exactly $b$ coefficients that are descendants of $i$ are chosen
  along with the coefficients of $S$.} The algorithm is given in Figure~\ref{fig:one}(b). Clearly the number of entries in the array $E[]$ is $Bn$ times $2^r$
where $r$ is the maximum number of ancestors of any node. It is easy
to see that $r=\log n +1$ and thus the number of entries is $n^2B$.
For $\ell_1$ measure we need to spend $O(B)$ time in the minimization
giving a running time of $O(n^2B^2)$. For $\ell_\infty$, we may
perform binary search and only need $\log B$ time (see \cite{GK04}).

\subsection{A Simple Improvement}

\begin{observation}
  A node $i$ al level $t_i$ can have at most $2^{t_i}-1$ descendants.
  Thus $E[i,b,S]$ is meaningful only for $2^{t_i}$ values of $b$
  (including $b=0$). Further, the number of nodes at level $t_i$ is
  $\lceil\frac{n}{2^{t_i}}\rceil$  and the number of possible subsets of 
  ancestors of a node is $2^{\log n + 1 - t_i}$.
\end{observation}

Thus the number of $E[]$ entries to fill corresponding to $i$ is $
2^{\log n + 1 - t_i}\min \{ B, 2^{t_i} \} $. The time takes is
$2^{\log n + 1 - t_i}\min \{ B^2, 2^{2t_i} \}$. Thus one way of
computing the total time taken is
\begin{eqnarray*}
& & \sum_{t_i=1}^{\log n} \frac{n}{2^{t_i}} 2^{\log n + 1 - t_i}\min \{ B^2, 2^{2t_i} \} + B^2 = \sum_{t_i=1}^{\log B} \frac{n}{2^{t_i}} 2^{\log n + 1 - t_i} 2^{2t_i} 
+ \sum_{t_i=\log B+1}^{\log n} \frac{n}{2^{t_i}} 2^{\log n + 1 - t_i} B^2 
 + B^2 \\
& & = \sum_{t_i=1}^{\log B} 2n^2 + \sum_{u=1}^{\log n - \log B} \frac{n}{2^{u+\log B}} 2^{\log n + 1 - u - \log B} B^2 
 + B^2  = \ 2n^2 \log B + 2n^2 \sum_{u=1}^{\log n - \log B} \frac{1}{4^u} + B^2 
\end{eqnarray*}

\noindent which is $O(n^2\log B)$. In case of $\ell_\infty$, the expression
$\sum_{t_i=1}^{\log n} \frac{n}{2^{t_i}} 2^{\log n + 1 - t_i}\min \{
B\log B, t_i2^{t_i} \} + B^2$ can be shown to be $O(n^2)$ using the
same scheme and change of variables as above.
%Observe that the above also reduces the total space complexity. But we
%omit the discussion since we will achieve a better space bound
%anyways.

\subsection{The Intuition and the new algorithm}
The properties that stands out from the above dynamic program are
\begin{itemize}\parskip=-0.05in
\item {\em There is no connection between $E[i,b,S]$ and $E[i,b',S']$ as long as $S \neq S'$.}
\item {\em We do not need $E[i,b,S]$ while computing $E[i',b',S']$ unless $i$ is a child of $i'$ and either  $S=S'$ or $S=S'\cup\{i'\}$.}
\item {\em And finally, there is no need to allocate space for $E[i,b,S]$ while computing $E[i',b',S']$ if $i$ is an ancestor (not a descendant) of $i'$.}
\end{itemize} 

The simplest view of the new algorithm that computes the same table
({\em but it is not stored in entirety at any time}) is a parallel
algorithm, where there is a processor at each node of the error tree.
The algorithm at a node $i$ with children $i_L,i_R$ can be described as follows:
\begin{enumerate}\parskip=-0.05in
\item The node $i$ receives $S$ from its parent and seeks to return an 
array of size $B$ (or less) corresponding to $E[i,b,S]$ for $0\leq b \leq B$.
It actually receives
\[ v(i,S)=\sum_{i'\in S,i \mbox{ left descendant of }i'} c_{i'} - 
\sum_{i'\in S,i \mbox{ right descendant of }i'} c_{i'} \]

\item To evaluate $\min_{b'} E[i_L,b',S\cup\{i\}]+ E[i_R,b-1-b',S\cup\{i\}] $ 
the node $i$ passes $S\cup\{i\}$ to both of its children, i.e., 
$v({i_L},S\cup\{i\})$ and $v(i_R,S\cup\{i\})$. 
The children return the two arrays of size $B$ (or less), 
and the $\min_{b'}$ is performed for each $b$. Note that the right child can reuse the same space needed by the left child.
\item Now $i$ passes $S$ to the children and asks for $E[i_L,b,S]$ for all $b$ and likewise for $i_R$.
\item The node $i$ can now compute all $E[i,b,S]$. The entire time spent at this node is $\min \{ 2^{2t_i},B^2\}$.
\item If $i$ is the overall root, then $i$ also performs a minimization over all $b$ to find the solution with {\bf at most $\mathbf B$} coefficients. 
\end{enumerate}

\begin{lemma} No node receives the value $v(i,S)$ twice for the same set $S$.
\end{lemma}
The above shows that the algorithm is correct and runs in time $O(n^2\log n)$ (and $O(n^2)$ for $\ell_\infty$).
The next lemma is also immediate from the description of the algorithm:
\begin{lemma}
The space required at node $i$ is $\min \{ B , 2^{t_i} \}$, since this space is used for all $S$.
\end{lemma}
Thus the total space required is $O(B \log (n/B))$ (the last $\log B$ levels use geometrically decreasing space which sums to $O(B)$ and $\log n - \log B = \log (n/B)$).
Therefore if we consider the algorithm that simulates the parallel algorithm, we can
conclude with
\begin{theorem}
We can compute the {\bf error} of optimum $B$ term wavelet synopsis in time $O(n^2\log B)$ (and $O(n^2)$ for $\ell_\infty$) using overall space $O(n+B \log (n/B))=O(n)$.
\end{theorem}
Observe that we can only compute the error, and we do not know which coefficients are in the synopsis.
\subsection{How do we find the coefficients?}
We now show how to retrieve the coefficients after finding the total
error.  When we find the optimum error, we also resolve (i) if the
topmost coefficient is present or not and (ii) what is the allocation
of the coefficients to the left and right children.  Armed with these
two pieces of information, {\em we simply recurse/recompute}, i.e., we
pass the appropriate set (or $v(i,S)$ values) to the two children and
their respective allocations. Each child now finds the total error
{\em restricted to its subtree} and each decides on the two pieces of
information to set up the recursive game.

\noindent{\bf Analysis:} Let the running time of the recompute strategy be $f(n)$. To find the optimum error, we spend $c n^2 \log B $ time and therefore we have the recursion:
\[ f(n) = c n^2 \log B + 2f(n/2) \]
If we unroll the recursion one step, we see that $f(n) = c n^2 \log B + 2c(n/2)^2 \log B + 4 f(n/4)$. We can immediately observe that we are setting up a geometric sum and we can bound $f(n)$ by $2cn^2 \log B$. Therefore we conclude:

\begin{theorem}
  We can compute the {\em complete solution, i.e., total error and the
    stored coefficients} of the optimum $B$ term wavelet synopsis in
  $O(n^2\log B)$ time ($O(n^2)$ for $\ell_\infty$) using overall space
  $O(n + B \log (n/B))$.
\end{theorem}

{\bf Caveat:} We have to be careful and ensure that when we output the
coefficients recursively, we output all the coefficients of the first
half before outputting all the coefficients of the next half. In the
process, we need to remember the partition of the buckets, the
parameter $b'$, for $\log n$ levels. But since we have to remember
only $1$ number, the total space is $O(n+B\log (n/B)+\log n)=O(n+B\log
(n/B))$.

\section{Unrestricted Wavelet Synopsis construction Algorithms} 
\label{sec:new}
We now show how to obtain an approximation algorithm for the
general/unrestricted wavelet synopsis construction problem. We focus
our attention on $\ell_k$ error, we indicate the changes necessary for
the weighted case appropriately. Recall that the Wavelet synopsis
problem is: Given a set of $n$ numbers $X=x_1,\ldots,x_n$, find a $Z
\in \real^n$ with at most $B$ non-zero entries such that $\| X -
\wai(Z) \|_k$ is minimized.

The following will be an important observation leading towards a
suitable algorithm: {\em If we observe the previous algorithm based on
  assigning a processor to each coefficient in the error tree, we
  immediately observe that if for different subsets of ancestors, we
  receive the same value, i.e., $v(i,S)=v(i,S')$ for $S'\neq S$, we
  need not redo the computation.}  {\bf Note:} that the savings cannot
be guaranteed and in order to achieve the savings we have to increase
the space bound.

\paragraph{Overview:} The above will form a kernel of our algorithm for the
(unrestricted) wavelet synopsis construction problem. We would
actually perform the computation {\em for all possible, anticipated
  values of $v(i,S)$. However, non-zero elements of $Z$ can have any
  real value and it is not clear how to restrict the set of values.}

In what follows, we first describe the algorithm assuming that the
wavelet coefficients belong to a set of anticipated values $R$.
Subsequently we describe how to
determine $R$ and more importantly, bound $|R|$.

\subsection{The Algorithm}
\begin{defn}
Let $E[i,v,b]$ be the minimum possible contribution to the overall 
error from all descendants of $i$ using exactly $b$ coefficients, under the 
assumption that the combined value of all ancestors chosen is $v$. 
\end{defn}

The overall answer is clearly $\min_b E[root,0,b]$. A natural dynamic
program is immediate, to compute $E[i,v,b]$ if we decide the best
choice is to allocate $b'$ coefficients to the left and let the
$i^{th}$ coefficient be $r$, then we need to add $E[i_L,v+r,b']$ and
$E[i_R,b-b'-1,v-r]$. The overall algorithm is:

\begin{enumerate}\parskip=-0.05in
\item The number of $b$ that are relevant to $i$ is $\min\{ B,2^{r_i} \}$. 
The node receives the $E[i_L,v',b'],E[i_R,v'',b'']$ from its children.
\item A non-root node computes $E[i,v,b]$ as follows:
\vspace{-0.05in}
\[ E[i,v,b] = \min \left \{ \begin{array}{ll}
\min_{r,b'} E[i_L,v+r,b'] + E[i_R,v-r,b-b'-1] & \mbox{~~~$i^{th}$ coefficient is $r$} \\ 
\min_{b'} E[i_L,v,b'] + E[i_R,v,b-b'] & \mbox{~~~$i^{th}$ coefficient not chosen} 
\end{array} \right.
\] 
\item  If $i$ is the root, then $i$ computes
\vspace{-0.05in} \[
\min_b \left \{ \begin{array}{ll}
\min_{r,b'} E[i_L,r,b'] + E[i_R,r,b-b'-1] & \mbox{~~~root coefficient is $r$}\\
\min_{b'} E[i_L,0,b'] + E[i_R,0,b-b'] & \mbox{~~~root coefficient not chosen} 
\end{array} \right.
\]
\end{enumerate}

Note that the root can figure out (i) the optimum error (ii) if any
coefficient corresponding to it is chosen and (iii) the value $r$ of
the coefficient. After the final solution is computed, we apply the
recompute strategy, and each node in the tree finds out if it has a
coefficient in the answer and its value. The running time is 
\[ \sum_i |R| \min \{ 2^{r_i},B \} \cdot
|R| \min \{ 2^{r_i},B \} = \sum_{t} |R|^2 \frac{n}{2^t} \min \{ 2^{2t},B^2 \} = |R|^2 nB
\]
\vspace{-0.1in}

For $\ell_\infty$ the bound is $\sum_{t} |R|^2 \frac{n}{2^t} \min \{
t2^{t},B \log B \} = O(n|R|^2\log^2 B)$. The required space can be
shown to be $O( RB\log (n/B))$ ensuring that the computation resembles a
post-order traversal of the tree and we do not the tables of the
children nodes once we are done. Thus for each level we may need at
most $2$ tables of size $R \min\{B,2^\ell\}$, which sums to the above..

\subsection{Computing $R$}
\begin{lemma}
\label{poo}
If the $\max_i |x_i|$ is $M$ then $\max_i |\wa(X)_i| \leq M$.
\end{lemma}
\begin{proof}
  The $1^{st}$ coefficient is the average of all values and therefore
  cannot exceed $M$.  Every other coefficient is half the average value of
  left half (of the support) minus half the average value of right half.
  Each cannot be more than $M$ in absolute value.
\end{proof}
\begin{lemma}
\label{boo}
  If the optimum solution is $Z^*$ then $\max_i |Z^*_i| \leq 2n^{\frac1k}M$.
\end{lemma}
\begin{proof}
  If $\max_i |\wai(Z^*)_i| \geq 2n^{\frac1k}M$ then $\|X - \wai(Z^*)\|_k \geq \|\wai(Z^*)\|_k - \|X\|_k $ and 
\[ \|\wai(Z^*)\|_k - \|X\|_k \geq  \|\wai(Z^*)\|_k - Mn^{\frac1k} \geq \max_i |\wai(Z^*)_i| - Mn^{\frac1k} \geq Mn^{\frac1k} \geq \|X\|_k\]
The all zero solution is a better solution, which is a contradiction. 
Now we apply Lemma~\ref{poo} and get $\max_i
  |\wa(\wai(Z^*))_i| = \max_i |Z^*_i| \leq 2n^{\frac1k}M$, which proves the lemma.
\end{proof}
In case of weighted $\ell_k$ the above is modified to $\max_i |Z^*_i|
\leq 2n^{\frac1k}M \frac 1 {\min_i \pi_i}$.
The next lemma follows from triangle inequality.
\begin{lemma}
If we round each non-zero value of the optimum $Z^*$ to the nearest multiple 
of $\delta$ thereby obtaining $\hat{Z}$, then $\| X - \wai(\hat{Z})\|_k \leq  \| X - \wai(Z^*)\|_k + \delta n^{\frac1k}$ and $|R| \leq \frac{2n^{\frac1k}M}{\delta}$.

\end{lemma}
  Therefore if we set $\delta = \epsilon M/n^{\frac1k}$ we can say that we have an additive approximation of $\epsilon M$ as well as $|R|  = O(\epsilon n^{\frac2k})$.
Therefore we conclude the following:
\begin{theorem}
  We can solve the Wavelet Synopsis Construction problem with $\ell_k$
  error with an additive approximation of $\epsilon M$ where $M=\max_i
  |x_i|$ in time $O(n^{1+\frac4k}B\epsilon^{-2})$ and space
  $O(n+n^{\frac1k}\epsilon^{-1} B \log (n/B))$. For $\ell_\infty$ the running
  time is $O(n\epsilon^{-2}\log^2 B)$.
\end{theorem} 

\section{The theme of space efficiency and applications}

A natural paradigm emerges from inspecting the above: 
{\em If we can
compute the total error and the best way to partition the problem into
two halves of $\frac{n}2$ elements, we do not need to store the entire
dynamic programming table} -- {\em and thereby save space.} 
If we can compute the
overall error in time $f(n)=An^\alpha$ where $A$ is independent of
$n$, then the time taken by the {\em Recompute} strategy is
$g(n)=f(n)+2g(n/2)$. The solution to the recurrence is 
$O(An^\alpha)$ if $\alpha>1$ and $O(An\log n)$ if $\alpha=1$.

We demonstrate the above idea in two examples. First, we show its
impact in space efficient V-Opt histogram construction.  Second, we
show the applicability in a new synopsis technique, {\em Extended
  Wavelets}.

The idea also improves several results on range query histograms --
however those algorithms are quite similar in spirit to the V-Opt
histogram construction and we relegate the discussion to a fuller
version of the paper. However the idea does help in reducing the space
bound across the board -- in fact for a large variety of problems it
is immediate that the notion of {\em working space}, the space
necessary to compute the {\em value} of the final answer, is not
required any more. We can compute the entire answer, in the
aforementioned working space.

\subsection{V-Opt Histograms}
\label{sec:hist}
The V-Opt histogram is a classic problem in synopsis construction.
Given a set of $n$ numbers $X=x_1,\ldots,x_n$ the problem seeks to
construct a $B$ piecewise constant representation $H$ such that $\|X -
H \|_2$ (or its square) is minimized.  Since their introduction in
query optimization in \cite{I93}, and subsequently in
approximate query answering (\cite{Aqua}, among others), histograms
have accumulated a rich history \cite{I03}.  Several different
optimization criteria have been proposed for histogram construction,
e.g., $\ell_1$, relative error, $\ell_\infty$, to name a few.  However
most of them are based on a dynamic program similar to the V-Opt case.
Thus the V-Opt histograms provide an excellent foil to discuss all of
the measures at the same time.  As mentioned in the introduction,
\cite{Jag98} gave a $O(n^2B)$ time algorithm to find
the optimum histogram using $O(nB)$ space. They observed that the
space could be reduced to $O(n)$ at the expense of increasing the
running time to $O(n^2B^2)$.  The data stream algorithms\footnote{Note
  that by the streaming model we refer to the ``sorted'' or
  ``aggregate'' model, most useful in time series data, where the
  input is $x_i$ in increasing order of $i$. Only \cite{GGIKMS02}
  applies to the general ``turnstile'' or ``update'' model, but seems
  to have high polynomial dependence on $B\epsilon^{-1}\log n$. See
  \cite{muthu-survey,pods02} for more details on data stream models.}
of \cite{GKS01} (extended in \cite{GKS04}) represent sparse dynamic
tables -- but the space is still $\tilde{O}(B^2)$, a quadratic in $B$.
In a those algorithms the $\tilde{O}(B^2)$ space performs a double role
of storing the coefficients as well as maintaining a frontier.  

This is somewhat remedied in \cite{GIMS02,MS04}, where a robust
wavelet representation of $\tilde{O}(B)$ coefficients is constructed 
and then a dynamic program in the fashion of \cite{Jag98} or \cite{GKS01} 
restricted to the {\em endpoints of the support regions} is used.  
The dynamic program of \cite{Jag98} can be used to compute the answer 
in $\tilde{O}(B)$ space, but with an extra factor of $B$ in
running time. Therefore, irrespective of offline or streaming computation
there was a tradeoff between large space and an increased running time
-- this is {\em the penalty} referred to in the introduction.
This is the first paper which removes that penalty and gives an algorithm that 
simultaneously achieves the best known space and time bounds.

\begin{center}
{\small
\begin{tabular}{|c|c|c|c|c|c|}
\hline
Paper & Stream & Factor & Time & Space &  Working space  \\
\hline
\hline
\cite{Jag98} & No & Opt & $O(n^2B)$ & $O(nB)$ & $O(n)$ \\ 
& &  & $O(n^2B^2)$ & $O(n)$ & $O(n)$ \\
\hline
\cite{GKS01} & Yes & $(1+\epsilon)$ & $ O(nB^2\epsilon^{-1}\log n $ & $O(B^2\epsilon^{-1} \log n)$ & --\\
\hline \cite{GIMS02} & Yes & $(1+\epsilon)$ & $O(n+B^3 \epsilon^{-8} \log^4 n)$ & $O(B^2\epsilon^{-4} \log^2 n)$ & -- \\  
& & &  $O(n+B^4 \epsilon^{-8} \log^4 n)$ & $O(B \epsilon^{-4} \log^2 n)$ & --\\ 
\hline
\cite{newver}& No & $(1+\epsilon)$ & $O(n + B^3(\epsilon^{-2} + \log n) \log n)$ & $O(n + B^2\epsilon^{-1})$ & $O(n + B\epsilon^{-1})$ \\ 
& Yes &  & $O(n+ (n/M)B^3 \epsilon^{-2} \log^3 n)$ & $O(M + B^2\epsilon^{-1} \log n) $ & -- \\ 
\hline \cite{MS04} & Yes & $(1+\epsilon)$ & $O(n+B^3 \epsilon^{-3} (\log 1/\epsilon) \log n)$ & $O(B\epsilon^{-2} (\log 1/\epsilon) \log n + B^2/\epsilon)$ & -- \\  
& & &  $O(n+B^4 \epsilon^{-3} (\log 1/\epsilon) \log n)$ & $O(B \epsilon^{-2} (\log 1/\epsilon) \log n + B/\epsilon)$ & --\\  
\hline
\hline 
{\bf This Paper} & No & Opt & $O(n^2B)$ & $O(n)$ & $O(n)$ \\
& No &  $(1+\epsilon)$ & $O(n + B^3(\epsilon^{-2} + \log n) \log n)$ & $O(n + B\epsilon^{-1})$ & $O(n + B\epsilon^{-1})$ \\
& Yes & $(1+\epsilon)$ & $O(n+B^3 \epsilon^{-3} (\log 1/\epsilon)\log n)$ & $O(B\epsilon^{-2} (\log 1/\epsilon) \log n + B/\epsilon)$ & -- \\
\hline
\end{tabular}
}
\end{center}

\paragraph{Algorithm idea:} Due to lack of space, we indicate the 
modification to the optimum algorithm. The modifications to the 
approximation and streaming algorithms are similar. The optimal 
algorithm maintains $E[i,b]$ which is the
minimum error of expressing the interval $[1,i]$ by at most $b$
buckets (intervals where the representation is constant). 
A natural dynamic programming arises: $E[i,b] = \min_{j<i}
E[j,b-1] + e(j+1,i)$ where $e(j,i)$ is the minimum error of a single
bucket\footnote{It is straightforward to show that the minimum error
  is achieved by the mean of $x_{j+1},\ldots,x_i$.}.  The running time
is $O(n^2B)$. If we are interested in computing only the final answer,
there is an $O(n)$ space algorithm which computes $E[i,1]$ for all $i$, and then
extends that to $b=2,3,$ etc.

If $i>\frac n 2$ we maintain $A[i]$ to be the starting point of the 
bucket that contains the $x_\frac{n}2$ for the best representation of $[1,i]$ 
by $b$ buckets, and $B[i]$ to be the ending point
of that interval, and $C[i]$ to be the number of buckets used before $A[i]$. 
This requires $O(n)$ space, and is updated as shown below. Now, after we compute 
$E[n,B]$ we can divide the problem into two parts, representing $[1,A[i]]$ using 
$C[i]$ buckets and $[B[i]+1,n]$ by $B - C[i] - 1$ buckets. {\em Note that each subproblem is defined on $\frac{n}{2}$ or less elements}. Therefore the {\em Recompute strategy} will run in time $O(n^2B)$ as well and compute all the coefficients.

\begin{figure}[htbp]
\begin{center}
\framebox{\small
\begin{minipage}{5.5in}
\begin{tabbing}11111\=111\=111\=111\=111\=111\=111\=111\=111\=111\=111\kill 
\bcc \> $A[i]=0$ if $i\leq \frac{n}2$ and $1$ otherwise. $B[i]=0$ if
  $i\leq \frac{n}2$ and $i$ otherwise. $c[i]=0$ for all $i$.\\
\icc \> For $b=2$ to $B$ do \\
\icc \> \> For $i=2$ to $n/2$ do \\
\icc \> \> \> $E[i,b]=\min_{j<i} E[j,b-1] + e(j+1,i)$ \\
\icc \> \> For $i=n/2$ to $n$ do \\
\icc \> \> \> $E[i,b]=\min_{j<i} E[j,b-1] + e(j+1,i)$ \\
\icc \> \> \> If $j$ (which achieved the minimum) $\leq \frac{n}2$ then $newA[i]=j+1,newC[i]=b,newB[i]=i$.\\
\icc \> \> \> else $newA[i]=A[j],newB[i]=B[j],newC[i]=C[j]$; \\
\icc \> \> $A \leftarrow newA, B \leftarrow newB, C \leftarrow newC$. \\
\icc \> Recurse using $A[n],B[n],C[n]$ to compute the coefficients.
\end{tabbing}
\end{minipage}
}
\end{center}
\vspace{-0.2in}
\caption{The $O(n)$ space optimum algorithm\label{fig:opt}}
\end{figure}

Observe that we wave kept the $E[j,b-1],E[i,b]$ notation, but we can
reuse two arrays of size $n$ for this purpose (and keep switching them
as $newE,E$ etc.) -- the overall space required is $O(n)$.  We now
know the final solution $E[n,B]$ and how to partition the problem.
For {\em offline approximation algorithm}, when we recurse, we have to
add the approximate error $E'[B[i]+1,C[i]+1]$ to all the elements on
the right subproblem (since we build histograms with error increasing by $1+\epsilon$ factor, this ``shift'' is needed). Due to lack of space, the details are relegated
to the full version.

\subsection{Extended Wavelets}
\label{sec:ext}
Extended wavelets were introduced in
\cite{DR03}. The central idea is that in case of multi-dimensional
data, there can be significant saving of space if we use a
non-standard way of storing the information. There are several
standard ways of extending 1-dimensional (Haar) wavelets to multiple
dimensions. The wavelet basis corresponds to high-dimensional squares.
But irrespective of the number of dimensions, the format of the
synopsis is a pair of numbers {\em (coefficient index,value)}.
In Extended Wavelets we perform wavelet decomposition independently in
each dimension but then 
%If we wish to store the coefficient $i$ in all the
%dimensions, we can store $i$ followed by a list of the values. This
%would use roughly half the space to store each coefficient (since we
%are storing 1 number per dimension). 
we store tuples consisting of 
the coefficient index, a bitmap indicating the dimensions for
which the coefficient in that dimension is chosen,and a list of
values. Since the coefficient number and the bitmap is shared across
the coefficients, we can store more coefficients than a simple union
of unidimensional transforms.

Notice that there is no interaction between the benefits of storing
coefficient $i$ and $i'$. The problem reduces naturally to a {\em
  Knapsack} problem with a twist that each item (coefficient $i$) can
be present in varying sizes (how many values corresponding to
different dimensions are stored). However the variant also has a
simplifying feature that the space bound is polynomially bounded,
therefore allowing a simple dynamic program. The program estimates
$E[i,b]$ which indicates the minimum error on using {\em at most} $b$ 
space and storing only a subset of the first $i$ coefficients.

The idea is relatively new, and it remains to be seen if Extended
wavelets are applied widely. But it is an intriguing and novel idea in
synopsis construction and serve as an example of the broad
applicability of the ideas in this paper. 
This paper is also the first (almost) linear ($O(B)$, ignoring $M$) space
algorithm in the streaming (as well as offline) model.  We present the
results on the optimum algorithms below\footnote{The input is $n$
  tuples in $M$ dimensions and the total synopsis size is $B$.  The
  papers \cite{DR03,GKS04} contain other approximation algorithms that
  are not relevant to our context. The extended version of \cite{GKS04} reduces
  Extended Wavelets to a problem similar to V-Opt histogram
  construction and gives a $O(NM)$ time algorithm using dynamic
  programming.  The ideas of this paper naturally implies improvements
  to the space requirement under the assumption that $B \ll NM$.  The
  reduction is somewhat detailed and is omitted in this draft.}.

\begin{center}
{\small
\begin{tabular}{|c|c|c|c|c|c|}
\hline
Paper & Stream &  Time & Space & Working Space \\
\hline
\hline
\cite{DR03} & No & $ O(nMB) $ & $O(nMB)$ & $O(nM+MB)$\\
\hline
 \cite{GKS04} & Yes & $O(nMB)$ & $O(MB+B^2)$ & $O(MB+B^2)$ \\
& & $O(nM \log M +B^2M^2)$ & $O(MB+B^2)$ & $O(MB+B^2)$ \\
 \hline
\hline
{\bf This Paper} & Yes & $O(nM \log M + B^2 \log M )$ & $O(MB)$ & $O(MB)$\\
\hline
\end{tabular} 
}
\end{center}

\paragraph{Algorithm Idea:} We follow the previous algorithms and introduce a 
few small changes and a more careful analysis. For each item $i$ we
compute the best profit if $i$ is allocated size $j$. This is done in
time $O(nM\log M)$ as in \cite{GKS04}. For each $1\leq j\leq M$ we
maintain the top $B/j$ items corresponding to size $j$. For each $j$ we
can achieve this in $O(B/j)$ space and $O(n)$ running time (using
details from \cite{GIMS02}), using overall $O(nM)$ time and $\sum_j
(B/j)j = O(BM)$ space. The optimum answer uses items and sizes from
this list only. The total number of item-size pairs are $\sum_j (B/j)
= O(B \log M)$.
 
We can sort this list in lexicographic order. %using time $O(B(\log M) \log B)$.
Suppose item $i$ has $x_i \geq 1$ occurrences (thus $\sum_i x_i =O(B \log M)$).
The dynamic program to extend the answer to $i$ (from the item before
$i$) first needs to guess/choose which of the $x_i$ occurrences
are used (or none) and compute the best solution for each $B$. The
time taken is $c(x_i+1)B$ at $i$, which totals to at most $2cB^2\log M$.

We maintain a $O(B)$ array where $P[z]$ corresponds to the best
profit for space $z$ up to the current $i$.  For {\em space
  efficiency}, for $z\geq B/2$ we keep track of $Q[z]$ which contains
the pair $\langle,i',r,b'\rangle$ s.t. the optimum solution for space
$z$ for current $i$ uses space $b' < B/2$ Upton $i'$ and a size $r$
copy of $i'$ with $b'+r\geq B/2$. In other words, the {\em crossing
  point} where we crossed $B/2$ space for that solution (which remains
same even if we extend it later).

We now recurse with $b,b' \leq B/2$ on the two parts. Now each item
contributes $c(x_i+1)B/2$ adding up to less than $cB^2 \log M$. Once
again we have a geometric sum which sums up to $O(B^2\log M)$ for the
entire recursion.~\\

\noindent {\bf Acknowledgments:} We would like to thank Hyoungmin 
Park and Kyuseok Shim for many interesting discussions.
{\small
\bibliographystyle{plain}

\begin{thebibliography}{10}

\bibitem{Aqua}
S.~Acharya, P.~Gibbons, V.~Poosala, and S.~Ramaswamy.
\newblock {The Aqua Approximate Query Answering System}.
\newblock {\em Proc. of ACM SIGMOD}, 1999.

\bibitem{pods02}
B.~Babcock, S.~Babu, M.~Datar, R.~Motwani, and J.~Widom.
\newblock Models and issues in data stream systems.
\newblock {\em PODS}, pages 1--16, 2002.

\bibitem{DR03}
A.~Deligiannakis and N.~Roussopoulos.
\newblock Extended wavelets for multiple measures.
\newblock In {\em SIGMOD Conference}, 2003.

\bibitem{GK04}
M.~Garofalakis and A.~Kumar.
\newblock Deterministic wavelet thresholding for maximum error metric.
\newblock {\em Proc. of PODS}, 2004.

\bibitem{GG02}
M.~N. Garofalakis and P.~B. Gibbons.
\newblock Wavelet synopses with error guarantees.
\newblock In {\em Proc. of ACM SIGMOD}, 2002.

\bibitem{GGIKMS02}
A.~C. Gilbert, S.~Guha, P.~Indyk, Y.~Kotidis, S.~Muthukrishnan, and Martin
  Strauss.
\newblock Fast, small-space algorithms for approximate histogram maintenance.
\newblock In {\em Proc. of ACM STOC}, 2002.

\bibitem{GIMS02}
S.~Guha, P.~Indyk, S.~Muthukrishnan, and M.~Strauss.
\newblock Histogramming data streams with fast per-item processing.
\newblock In {\em Proc. of ICALP}, 2002.

\bibitem{GKS04}
S.~Guha, C.~Kim, and K.~Shim.
\newblock {XWAVE}: Optimal and approximate extended wavelets for streaming
  data.
\newblock {\em Proceedings of VLDB Conference}, 2004.

\bibitem{GKS01}
S.~Guha, N.~Koudas, and K.~Shim.
\newblock {Data Streams and Histograms}.
\newblock In {\em Proc. of STOC}, 2001.

\bibitem{newver}
S.~Guha, N~Koudas, and K.~Shim.
\newblock Approximation algorithms for histogram construction problems.
\newblock {\em Technical Report, the full version of \cite{GKS01}, available at
  http://www.cis.upenn.edu/~sudipto/mypapers/histjour.pdf.gz}, 2004.

\bibitem{I93}
Y.~E. Ioannidis.
\newblock Universality of serial histograms.
\newblock In {\em Proc. of the VLDB Conference}, 1993.

\bibitem{I03}
Y.~E. Ioannidis.
\newblock The history of histograms (abridged).
\newblock {\em Proc. of VLDB Conference}, pages 19--30, 2003.

\bibitem{Jag98}
H.~V Jagadish, N.~Koudas, S.~Muthukrishnan, V.~Poosala, K.~C. Sevcik, and
  T.~Suel.
\newblock {Optimal Histograms with Quality Guarantees}.
\newblock In {\em Proc. of the VLDB Conference}, 1998.

\bibitem{yossi1}
Y.~Matias and D.~Urieli.
\newblock Optimal workload-based wavelet synopses.
\newblock {\em TR-TAU}, 2004.

\bibitem{MVW98}
Y.~Matias, J.~Scott Vitter, and M.~Wang.
\newblock { Wavelet-Based Histograms for Selectivity Estimation}.
\newblock {\em Proc. of ACM SIGMOD}, 1998.

\bibitem{muthu-survey}
S.~Muthukrishnan.
\newblock Data streams: Algorithms and applications.
\newblock {\em Survey available on request at {\tt muthu@research.att.com}},
  2003.

\bibitem{muthu-wave}
S.~Muthukrishnan.
\newblock Workload optimal wavelet synopsis.
\newblock {\em DIMACS TR}, 2004.

\bibitem{MS04}
S.~Muthukrishnan and M.~Strauss.
\newblock Approximate histogram and wavelet summaries of streaming data.
\newblock {\em DIMACS TR 52}, 2003.

\end{thebibliography}

}
\end{document}